\documentclass[12pt,preprint]{aastex}
\usepackage{apjfonts}
\usepackage{epstopdf}
\usepackage{graphicx}
\usepackage{lineno}
\usepackage{ulem}
\usepackage{lscape}
\usepackage{pdflscape}
\usepackage[top=2.54cm, bottom=0.2cm, left=3.18cm,right=3.18cm]{geometry}

\begin{document}

\title{A catalogue of double-mode high amplitude $\delta$ Scuti stars in the Galaxy and their statistical properties}
\shorttitle{Catalogue of Double-Mode HADS stars }

\author{Tao-Zhi Yang\altaffilmark{1},
Zhao-Yu Zuo\altaffilmark{1,*},
Xin-Yue Wang\altaffilmark{2},
Xiao-Ya Sun\altaffilmark{1},
Rui-Xuan Tang\altaffilmark{1}
}

\altaffiltext{1}{
Ministry of Education Key Laboratory for Nonequilibrium Synthesis and Modulation of Condensed Matter, School of physics, Xi'an Jiaotong University, Xi'an 710049, China; *E-mail: zuozyu@xjtu.edu.cn (ZYZ);}

\altaffiltext{2}{
School of Management, Xi'an Jiaotong University, 710049 Xi'an, People's Republic of China;  }


\begin{abstract}
We present the first catalogue of double-mode and multi-mode high amplitude $\delta$ Scuti star (HADS) in the Galaxy. 
The catalogue contains source name, coordinates, radial modes such as the fundamental (F, period P0), first-overtone (1O, period P1), 
second-overtone (2O, period P2), and the third-overtone (3O, period P3) if available, period ratios, magnitude, and the relevant literature. 
Totally, 155 sources were collected until March 2021, in which 142 HADS with double-mode (132 with F and 1O, and 10 with 1O and 2O), 
11 triple-mode, and 2 quadruple-mode. Statistical analysis shows clear features: P0 lies in a range of 0.05 days to 0.175 days 
(sample: 132 double-mode HADS pulsating in F and 1O); P1/P0 lies in a range of 0.761 $-$ 0.787 (sample: 142 with P0 and P1), in which 
about 90\% in 0.765 $-$ 0.785, which is wider than previous studies. The Petersen diagram was created with a much larger sample 
(144 HADS with P0 and P1) and we find that stars with periods in 0.05 $-$ 0.1 days scatter largely from the updated linear relation 
(i.e., Eq.~\ref{eq:linefit}), the reason of which however needs further investigation.
Particularly, we discover that the ratio P2/P1 (sample: 21 HADS with P1 and P2) equals 0.802$\pm$0.004, 
which could be viewed as a possible indicator to identify the modes 1O and 2O for multi-mode HADS.
In addition, several unusual stars are pointed out, which may need more attention to their pulsations and stellar parameters in the future.

\end{abstract}


\section{Introduction}

$\delta$ Scuti ($\delta$ Sct) stars are intermediate-mass pulsating stars with periods below 0.3 d and light amplitudes from mmag to 0.9 mag \citep{2013AJ....145..132C,2017ampm.book.....B}. They pulsate in radial as well as non-radial modes excited mainly in the $\kappa$ mechanism \citep{2000ASPC..210....3B,2010aste.book.....A,2014ApJ...796..118A,2020MNRAS.498.4272M}.  In the Hertzsprung-Russell (HR) diagram, these pulsating stars lie in the overlapping region of the main sequence (MS) and the classical instability strip, corresponding to spectral types from A0 to F6 and luminosity classess from III to V. They can be found at MS and post-MS, as well as pre-MS stage. The mass of $\delta$ Sct stars is usually metallicity-dependent, that's to say, Population I with mass of about 1.5 to 2.3 M$_{\odot}$ is higher than Population II of about 1.0 to 1.3 M$_{\odot}$ \citep{2011AJ....142..110M}. Usually, $\delta$ Sct stars are considered as the Population I stars of the flat Milk Way component in the young Galactic disk, while the Population II analogues are sometimes classified as SX Phoenecis stars (SX Phe) \citep{2011AJ....142..110M}. The majority of $\delta$ Sct stars pulsate in many non-radial modes and their amplitude spectra appear very messy, which challenges the mode identification.

Amongst the $\delta$ Sct stars, one species that pulsates in one or several radial modes with large amplitudes, is usually called as High-Amplitude $\delta$ Scuti stars (HADS). HADS are easily recognized by their large-amplitude of A$_{V}$ $\ge$ 0.3 mag and non-sinusoidal light curves from observations \citep{2000ASPC..210....3B}. Compared to $\delta$ Sct stars, HADS usually rotate with slow rotational velocities of $v$ sin $i$ $\le$ 40 km s$^{-1}$ \citep{2012A&A...537A.120Z,2015MNRAS.450.2764N,2021MNRAS.504.4039B}, which might be a precondition for their high amplitudes \citep{2000ASPC..210....3B}. In HR Diagram, ground-based observations indicate that HADS seem to occupy a narrow strip in the $\delta$ Sct instability region with a width of about 300 K in temperature, where the efficiency of mode excitation might be maximized \citep{1996A&A...312..463P,2000ASPC..210..373M}. However, photometric observations from space mission reveal that HADS could be spread beyond the narrow strip \citep{2016MNRAS.459.1097B}. Although some studies suggest the properties of HADS might be due to their evolution stage being in post-MS state \citep{1996A&A...312..463P,2017ampm.book.....B}, there is still no clear consensus on the physical difference between HADS and the lower amplitude $\delta$ Sct stars \citep{2019MNRAS.490.4040A}. Another subgroup, which is not easy to be distinguished from HADS, called "SX Phe". Typically, they have high spatial motion, low metallicities \citep{1980ApJ...235..153B,1989AJ.....97..431E,1989upsf.conf..215N,2017MNRAS.466.1290N,1970PASP...82..274E,1979ApJS...41..413E,2010AJ....139.2639L,2012AJ....144...92Y,2019PASP..131f4202Z}, and usually located in globular clusters and dwarf galaxies \citep{2000A&A...359..597R,2003AJ....125.3165J,2004AJ....128..287J,2003MNRAS.340.1205M,2010AJ....140..328G}. However field SX Phe stars were also found \citep[$\sim 50$,][]{2010A&A...515A..39F,2012MNRAS.426.2413B}, which is very similar to Population I HADS, in light variations and pulsation modes \citep{2010A&A...515A..39F}.  In this work, any confirmed SX Phe star was removed from the sample.

Generally, HADS mainly pulsate in one or two dominant frequencies which are radial modes \citep{2000ASPC..210....3B,2018RAA....18....2Y,2018ApJ...861...96X}. In fact, low-amplitude non-radial modes can also be found in some HADS, for example, AI Vel \citep{1992MNRAS.254...59W} and V974 Oph \citep{2003A&A...409.1031P}. For double-mode HADS, which pulsate in the first two radial modes, i.e. the radial fundamental mode (F, period P0) and first-overtone mode (1O, period P1), the AAVSO International Variable Star Index \citep[VSX\footnote{the AAVSO International Variable Star Index: {https://www.aavso.org/vsx/}},][]{2014yCat....102027W} designate them as HADS(B). The main advantage of studying HADS(B) is the simplification of mode identification through period ratios, as the ratio of P1/P0 of most HADS(B) lies in a narrow range of 0.77 $\le$ P1/P0 $\le$ 0.78 \citep{2000ASPC..210..373M}.

However for a long time, the sample of HADS is small, not to mention HADS(B), which hampered the study among HADS, that's to say, the relation between the single and double/multi-mode HADS and their stellar parameters \citep{2021AJ....161...27Y,2013AcA....63..379P}, the distribution of dominant periods and statistical properties of HADS(B), and so on. In recent years, with more ground-based surveys and space missions, e.g. Catalina Sky Survey (CSS, \citealt{2003DPS....35.3604L,2014ApJS..213....9D,2017MNRAS.469.3688D}), LINEAR \citep{2013AJ....146..101P}, RATS \citep{2011MNRAS.417..400R}, MACHO \citep{1997ApJ...486..697A,2000ApJ...536..798A}, ZTF \citep{2019PASP..131a8002B,2019PASP..131g8001G}, ASAS-SN \citep{2014AAS...22323603S,2020MNRAS.493.4186J}, OGLE survey \citep{2013AcA....63..379P,2020AcA....70..241P} and $Kepler$ mission \citep{2010Sci...327..977B,2010ApJ...713L..79K}, the number of HADS has increased dramatically \citep{2012MNRAS.424.2528S,2016JAVSO..44....6F}. So it is a very good opportunity to investigate the overall properties and stellar parameters of HADS(B), which may help us understand what determines the number of radial modes in which a star pulsates and explore if there is different physics between the single and double/multi-mode HADS.

In this paper, we collected HADS(B) and multi-mode HADS as far as possible, and presented the first catalogue for this kind of objects. The aim of this catalogue is to provide some basic information on HADS(B) and multi-mode HADS, including their coordinate, pulsating modes, period ratios, and the corresponding reference. However, no attempt has been made to compile complete reference lists and much effort has been made to avoid errors to the best of our knowlewdge. With this catalogue, a statistical analysis between the periods and their ratios was performed, and the relationship between them was further investigated. 

\section{Data collection}

The VSX is a public database, which aims to bring all the information on variable stars in new and/or ongoing surveys together. It provides a simple web interface for public to access, review and revise the metadata of variable stars \citep{2014yCat....102027W}.

In this work, we collected 152 HADS stars signed with HADS(B) stars from the database. Among them, five stars (i.e. MGAB-V1192, MGAB-V1182, MGAB-V1185, MGAB-V1163, MGAB-V1190) with only P0 (i.e., 0.05314, 0.05545, 0.06125, 0.04332, and 0.05598 d, respectively) are provided in the database. To verify their double-mode features, we re-analysised their pulsations using the public photometric data and found: MGAB-V1192 and MGAB-V1163 have no other periods; MGAB-V1185 has another period: P1=0.04233 d, but with a period ratio of 0.04233/0.06125=0.6911, not within the typical ratio of radial overtones to fundamental mode ; MGAB-V1182 has another period: P1=0.04333 d, with a period ratio of 0.04333/0.05545=0.78142; MGAB-V1190 has another period P1: 0.07205 d, if assumed it is the fundamental radial mode, then it has a period ratio of 0.05598/0.07205=0.77696, which is within the typical value of period ratio. Hence MGAB-V1182 and MGAB-V1190 are included in our sample. Another three stars: OGLE-GD-DSCT-0018, OGLE-GD-DSCT-0024, and OGLE-GD-DSCT-0036, are pulsators with multi-periods: (P0, P2, P3, no P1), (P0, P2, no P1) and (P1, P3, no P0), respectively. Due to the lack of P0 or P1, they are not included in the sample.

In addition, we also collected 9 stars, i.e., AE UMa \citep{2001AA...368..880P}, GSC 02583-00504 \citep{2003IBVS.5442....1W}, CzeV293 and CzeV581 \citep{2016NewA...46...85S}, KIC 5950759 \citep{2018ApJ...863..195Y}, KIC 10284901 \citep{2019ApJ...879...59Y}, ASASSN-V J061658.97-213318.9 \citep{2020MNRAS.493.4186J}, CSS J213934.3-050020 \citep{2021PASP..133e4201S}, and KIC 10975348 \citep{2021AJ....161...27Y} from recent literature. Among them, two stars, i.e., AE UMa and GSC 02583-00504, are already listed in VSX but classified as SXPHE(B), i.e., double-mode SX Phe stars. However, recent studies suggest they should belong to Population I HADS stars \citep{2017MNRAS.467.3122N,2003IBVS.5442....1W}. So they are also included in this sample.

In total, 155 stars were collected in this sample, among which each star possesses at least two radial pulsation modes. We note in the present sample, only Galactic HADS stars were taken into account, although HADS stars in other galaxies (Magellanic Clouds in particular) is large, which may need a separate study. For simplification, the radial fundamental mode and the first three radial overtones are abbreviated as F, 1O, 2O, and 3O in this work. With carefully inspection, we found there are 132 stars pulsating in F+1O, 10 in 1O+2O, 11 in triple-mode (10 stars in F+1O+2O, and 1 in 1O+2O+3O), and 2 stars in quadruple-mode (F+1O+2O+3O).

\begin{landscape}
\begin{deluxetable}{lccclcccll}
\tabletypesize{\small}
\tablewidth{0pc}
\tablenum{1}
\tablecaption{List of the double- and multi- mode HADS (Sorted by coordinate).\label{tab:List1}}
\tablehead{
\colhead{Name}   &
\colhead{R.A.(J2000)}      &
\colhead{Dec.(J2000)}      &
\colhead{P0 (day)} &
\colhead{P1 (day)} &
\colhead{P2 (day)} &
\colhead{P1/P0} &
\colhead{P2/P1} &
\colhead{Mag} &
\colhead{Reference}
}
\startdata

NSV 14800 & 00 01 16.22 & -60 36 57.1                 & 0.15784 & 0.12207 &- & 0.77339 &- & 9.70 - 10.20 V &   \citet{2005AA...440.1097P} \\
V0823 Cas & 00 05 42.38 & +63 24 14.2                 & 0.66900 & 0.51269 & 0.41092 & 0.76635 & 0.80150 & 10.80 - 11.37 V & \citet{2006AA...445..617J} \\
FASTT 8 & 00 39 09.42 & +00 40 12.1                   & 0.07302 & 0.05712 &- & 0.78223 &- & 13.7 - 14.1 V &   \citet{2009ApJ...696..870D} \\
CSS\_J021145.1+371038 & 02 11 45.13 & +37 10 38.3     & 0.10529 & 0.08151 &- & 0.77412 &- & 15.81 - 16.38 V & \citet{2018PZ.....38....1K} \\
GSC 03693-01705 & 02 12 19.83 & +57 00 16.4           & 0.09109 & 0.07047 &- & 0.77367 &- & 13.41 - 13.80 V & \citet{2004AJ....127.2436W} \\
RV Ari & 02 15 07.46 & +18 04 28.0                    & 0.09313 & 0.07195 &- & 0.77256 &- & 11.61 - 12.30 V &  \citet{2002AA...393..555P} \\
CSS\_J022438.9+262101 & 02 24 38.93 & +26 21 01.6     & 0.10158 & 0.08012 &- & 0.78871 &- & 13.36 - 13.83 V & \citet{2018PZ.....38....1K} \\
CSS\_J023722.9+382847 & 02 37 22.96 & +38 28 47.2     & 0.11068 & 0.08544 &- & 0.77199 &- & 14.40 - 14.98 V & \citet{2018PZ.....38....1K} \\
CSS\_J025539.8+314518 & 02 55 39.90 & +31 45 18.7     & 0.12890 & 0.09952 &- & 0.77205 &- & 16.20 - 16.78 V & \citet{2018PZ.....38....1K} \\
GSC 02860-01552 & 03 16 02.70 & +43 20 34.3           & 0.13831 & 0.10675 &- & 0.77182 &- & 12.52 - 13.02 V &  \citet{2010AA...520L..10B} \\
DDE 107 & 03 41 53.63 & -06 53 52.7                   & 0.08005 & 0.06199 &- & 0.77445 &- & 13.20 - 13.80 V & \citet{2009ApJ...696..870D} \\
CSS\_J034452.1+171634 & 03 44 52.19 & +17 16 34.6     & 0.07482 & 0.05784 &- & 0.77306 &- & 15.27 - 15.76 V & \citet{2018PZ.....38....1K}\\
V1384 Tau & 03 54 07.27 & +07 59 15.4                 & 0.13979 & 0.10739 &- & 0.76823 &- & 11.0 - 11.4 V &   \citet{2009PZP.....9...26K} \\
V1392 Tau & 04 26 05.90 & +01 26 26.2                 & 0.07443 & 0.05790 &- & 0.77795 &- & 12.00 - 12.55 V&  \citet{2013PZ.....33....6K}  \\
USNO-B1.0 1332-0126848 & 04 28 53.38 & +43 15 10.9    & 0.13715 & 0.10559 &- & 0.76988 &- & 14.4 - 15.0 V  & \citet{2018PZ.....38....1K}\\
USNO-B1.0 1329-0132547 & 04 44 37.78 & +42 54 34.4    & 0.16190 & 0.12413 &- & 0.76676 &- & 15.0 - 15.3  V &  \citet{2014PZ.....34....1K} \\
V542 Cam & 04 53 46.52 & +68 28 26.5                  & 0.17477 & 0.13400 &- & 0.76663 &- & 11.55 - 12.0 R &  \citet{2012PZP....12....6K} \\
CSS\_J051346.4+065924 & 05 13 46.48 & +06 59 24.2     & 0.10297 & 0.07972 &- & 0.77418 &- & 16.09 - 16.66 V & \citet{2018PZ.....38....1K} \\
GSC 02900-00317 & 05 16 25.51 & +41 03 56.8           & - & 0.16046  & 0.13064 & - & 0.81416 & 12.65 - 12.96 V & \citet{2018PZ.....38....1K}\\
GSC 04757-00461 & 05 23 54.48 & -03 07 32.3           & 0.13253 & 0.10194 &- & 0.76916 &- & 14.70 - 15.28 V & \citet{2009ApJ...696..870D} \\
CzeV581 & 05 46 27.21 & +31 11 09.9                   & 0.08814 & 0.06822 &- & 0.77398 &- & 15.35 V &       \citet{2016NewA...46...85S}\\
V803 Aur & 06 12 13.90 & +31 48 24.4                  & 0.07106 & 0.05503 & 0.04439 & 0.77442 & 0.80665 & 12.89 - 13.20 V & \citet{2014PZP....14....1K} \\
V2855 Ori & 06 15 17.73 & +06 04 12.6                 & 0.05808 & 0.04483 &- & 0.77182 &- & 9.93 - 10.36 V &   \citet{2005AA...440.1097P} \\
ASASSN-V J061658.97-213318.9 & 06 16 58.97 &-21 33 18.9 &   0.10183 & 0.07875 &- & 0.77336 &- & 13.27 - 13.67 V & \citet{2020MNRAS.493.4186J} \\
ASAS J062542+2206.4 & 06 25 41.61 & +22 06 19.5       & 0.15265 & 0.11731 &- & 0.76848 &- & 12.20 - 12.65 V & \citet{2015PZP....15....9K} \\
Brh V128 & 06 44 01.06 & +22 44 31.7                  & 0.15338 & 0.11770 &- & 0.76736 &- & 12.4 - 12.7 V  &  \citet{2004IBVS.5552....1B} \\
2MASS J06451725+4122158 & 06 45 17.25 & +41 22 15.9   & 0.05001 & 0.03869 & - & 0.77369 &- & 14.00 - 14.50 V & \citet{2010PASP..122...17J} \\
GSC 762-110   & 07 12 19.41 & +09 21 06.7             & 0.19451 & 0.14862 & 0.11908 & 0.76407 & 0.80124 & 10.54 - 10.73 V & \citet{2008AA...478..865W} \\
CSS\_J072643.6+413522 & 07 26 43.58 & +41 35 23.1     & 0.09032 & 0.07023 &- & 0.77758 &- & 16.71 - 17.41 V & \citet{2018PZ.....38....1K} \\
NSVS 7293918 & 07 44 38.60 & +29 12 22.8              & 0.08854 & 0.06850 &- & 0.77372 &- & 12.649 (0.313) R & \citet{2012IBVS.6015....1W} \\
AI Vel & 08 14 05.15 & -44 34 32.9                    & 0.11157 & 0.08621 &- & 0.77266 &- & 6.15 - 6.76 V &    \citet{1985AA...151..403B} \\
V733 Pup & 08 18 06.98 & -22 14 07.7                  & 0.22871 & 0.17423 &- & 0.76180 &- & 12.05 - 12.8 V & \citet{2011PZP....11....30K} \\
CSS\_J082237.3+030441 & 08 22 37.42 & +03 04 41.8     & 0.09055 & 0.07030 &- & 0.77636 &- & 13.31 - 13.83 V & \citet{2018PZ.....38....1K} \\
NT Cam & 08 24 17.52 & +74 30 25.4                    & 0.08242 & 0.06505 &- & 0.78925 &- & 12.74 - 13.13 V & \citet{2017PASP..129j4502K} \\
VZ Cnc & 08 40 52.12 & +09 49 27.2                    & - & 0.17836  & 0.14280 & - & 0.80063 &  7.18 -  7.91 V & \citet{1968Obs....88...58M}\\
CSS\_J085636.3-022534 & 08 56 36.42 & -02 25 35.2     & 0.08827 & 0.06839 &- & 0.77471 &- & 14.32 - 14.85 V & \citet{2018PZ.....38....1K} \\
LINEAR 5588339 & 08 58 54.72 & +15 22 09.7            & 0.05763 & 0.04460 &- & 0.77385 &- & 13.72 - 14.10 V & \citet{2009ApJ...696..870D} \\
V0526 Vel & 09 03 13.34 & -52 02 28.7                 & - & 0.07746  & 0.06245 & - & 0.80622 & 9.60 - 9.95 V & \citet{2011PZP....11....30K}\\
GSC 04135-00504 & 09 30 23.95 & +61 20 33.9           & 0.06150 & 0.04771 &- & 0.77581 &- & 12.40 - 12.90 V & \citet{2017PASP..129j4502K} \\
SSS\_J093025.5-215630 & 09 30 25.52 & -21 56 31.6     & 0.09585 & 0.07424 &- & 0.77449 &- & 12.67 - 13.21 V & \citet{2018PZ.....38....1K} \\
AE UMa        & 09 36 53.16 & +44 04 00.4             & 0.08602 & 0.06653 &- & 0.77343 &- & 10.86 - 11.52 V &  \citet{2001AA...368..880P} \\
ASAS J094303-1707.3 & 09 43 02.81 & -17 07 15.9       & 0.09918 & 0.07652 &- & 0.77153 &- & 11.75 - 12.12 V &  \citet{2005AA...440.1097P} \\
VX Hya & 09 45 46.85 & -12 00 14.3                    & 0.22339 & 0.17272 &- & 0.77318 &- & 10.21 - 10.96 V & \citet{1966ApJ...143..852F} \\
SSS\_J095011.1-244057 & 09 50 11.12 & -24 40 58.0     & 0.06839 & 0.05302 &- & 0.77525 &- & 15.18 - 15.50 V & \citet{2009ApJ...696..870D}\\
SSS\_J095657.2-231722 & 09 56 57.19 & -23 17 22.9     & 0.05667 & 0.04425 &- & 0.78083 &- & 16.2 - 16.74 V & \citet{2009ApJ...696..870D}\\
OGLE-GD-DSCT-0003 & 10 36 21.58 & -62 33 14.7         & 0.07517 & 0.05827 &- & 0.77515 &- & 18.33 - 18.53 I & \citet{2013AcA....63..379P} \\
OGLE-GD-DSCT-0007 & 10 41 08.32 & -61 42 16.9         & 0.11941 & 0.09222 &- & 0.77233 &- & 14.01 - 14.14 I & \citet{2013AcA....63..379P}\\
OGLE-GD-DSCT-0008 & 10 41 47.88 & -61 39 55.7         & 0.19804 & 0.15152 &- & 0.76509 &- & 13.27 - 13.41 I & \citet{2013AcA....63..379P} \\
OGLE-GD-DSCT-0010 & 10 42 42.80 & -61 40 59.7         & 0.11594 & 0.08934 &- & 0.77058 &- & 18.14 - 18.27 I & \citet{2013AcA....63..379P} \\
OGLE-GD-DSCT-0012 & 10 42 51.62 & -61 35 21.4         & 0.16748 & 0.13153 &- & 0.78537 &- & 13.62 - 13.75 I & \citet{2013AcA....63..379P} \\
OGLE-GD-DSCT-0014 & 10 43 44.31 & -61 30 41.2         & 0.07185 & 0.05567 &- & 0.77471 &- & 18.55 - 19.00 I & \citet{2013AcA....63..379P} \\
OGLE-GD-DSCT-0016 & 10 46 27.35 & -60 44 01.2         & 0.06743 & 0.05223 &- & 0.77464 &- & 16.09 - 16.36 I&  \citet{2013AcA....63..379P}  \\
OGLE-GD-DSCT-0020 & 10 48 57.42 & -61 36 34.1         & 0.21544 & 0.16404 &- & 0.76143 &- & 13.88 - 14.07 I & \citet{2013AcA....63..379P} \\
BPS BS 16553-0026 & 10 52 48.49 & +41 54 35.3         & 0.12551 & 0.09695 &- & 0.77248 &- & 12.75 - 12.95 R &  \citet{2010AA...520L..10B} \\
OGLE-GD-DSCT-0025$^{(a)}$ & 10 53 17.74 & -61 14 47.7 & 0.17406 & 0.13311 & 0.10646  & 0.76473 & 0.79983 & 13.16 - 13.29 I &\citet{2013AcA....63..379P}\\
V0474 UMa & 10 55 02.50 & +61 42 17.2                 & 0.06404 & 0.04961 &- & 0.77469 &- & 13.43 - 13.86 V & \citet{2014IBVS.6122....1W}\\
OGLE-GD-DSCT-0027 & 10 56 22.73 & -61 33 28.0         & 0.11764 & 0.09067 &- & 0.77072 &- & 16.04 - 16.16 I & \citet{2013AcA....63..379P}\\
OGLE-GD-DSCT-0028 & 10 56 25.25 & -61 49 38.1         & 0.06893 & 0.05338 &- & 0.77445 &- & 15.54 - 15.69 I & \citet{2013AcA....63..379P}\\
OGLE-GD-DSCT-0029 & 10 56 44.62 & -61 24 23.7         & 0.06143 & 0.04740 &- & 0.77155 &- & 16.52 - 16.65 I & \citet{2013AcA....63..379P} \\
OGLE-GD-DSCT-0030 & 10 56 52.71 & -60 41 36.1         & 0.12661 & 0.09754 &- & 0.77043 &- & 16.46 - 16.63 I & \citet{2013AcA....63..379P} \\
OGLE-GD-DSCT-0031 & 10 57 53.54 & -61 54 01.7         & 0.11676 & 0.09061 &- & 0.77606 &- & 18.55 - 18.93 I & \citet{2013AcA....63..379P} \\
OGLE-GD-DSCT-0032 & 10 58 26.38 & -61 14 42.0         & 0.08271 & 0.06464 &- & 0.78156 &- & 16.47 - 16.83 I & \citet{2013AcA....63..379P} \\
OGLE-GD-DSCT-0033 & 10 59 21.02 & -60 48 00.9         & 0.06331 & 0.04894 & 0.03949 & 0.77302 & 0.80691 & 15.88 - 15.99 I & \citet{2013AcA....63..379P}\\
OGLE-GD-DSCT-0037 & 11 09 23.01 & -60 51 20.8         & 0.23460 & 0.17878 &- & 0.76209 &- & 15.80 - 15.94 I & \citet{2013AcA....63..379P}\\
V899 Car & 11 09 52.24 & -60 57 56.7                  & 0.11080 & 0.08585 &- & 0.77482 &- & 11.8 - 12.4 V & \citet{2009PZP.....9...26K}\\
LINEAR 1683151 & 11 32 05.40 & -03 48 27.5            & 0.06185 & 0.04821 &- & 0.77949 &- & 16.24 (0.51) V&   \citet{2014PASP..126..509P} \\
OGLE-GD-DSCT-0040 & 11 32 10.16 & -60 42 44.6         & 0.09581 & 0.07406 &- & 0.77304 &- & 18.22 - 18.57 I & \citet{2013AcA....63..379P} \\
CSS\_J114944.1+185128 & 11 49 44.16 & +18 51 28.1     & 0.08968 & 0.07021 &- & 0.78289 &- & 16.89 - 17.60 V & \citet{2018PZ.....38....1K}\\
LINEAR 2653935 & 11 59 42.51 & +06 08 22.0            & 0.09521 & 0.07460 &- & 0.78357 &- & 16.52 (0.35) V &  \citet{2014PASP..126..509P} \\
CSS\_J120037.4+122805 & 12 00 37.39 & +12 28 05.3     & 0.09032 & 0.07086 &- & 0.78454 &- & 16.76 - 17.66 V & \citet{2018PZ.....38....1K} \\
GSC 07243-00871 & 12 08 49.77 & -36 33 11.1           & 0.06003 & 0.04648 &- & 0.77427 &- & 13.02 - 13.15 V &  \citet{2010AA...520L..10B} \\
SSS\_J130636.7-354622 & 13 06 36.82 & -35 46 23.0     & 0.13799 & 0.10639 &- & 0.77099 &- & 13.99 - 14.74 V & \citet{2018PZ.....38....1K} \\
OGLE-GD-DSCT-0044 & 13 15 58.51 & -64 37 25.7         & - & 0.15639  & 0.12483 & - & 0.79820 & 14.74 - 14.88 I & \citet{2013AcA....63..379P} \\
OGLE-GD-DSCT-0045 & 13 16 00.48 & -64 58 28.1         & 0.11781 & 0.09105 &- & 0.77281 &- & 14.96 - 15.12 I & \citet{2013AcA....63..379P}\\
OGLE-GD-DSCT-0048 & 13 24 14.87 & -65 09 03.9         & 0.10740 & 0.08293 & 0.06655 & 0.77216 & 0.80248 & 16.22 - 16.37 I& \citet{2013AcA....63..379P}\\
OGLE-GD-DSCT-0049 & 13 24 21.24 & -64 55 55.3         & 0.36142 & 0.27315 & 0.21699 & 0.75577 & 0.79440 & 14.43 - 14.70 I& \citet{2013AcA....63..379P} \\
OGLE-GD-DSCT-0071$^{(b)}$ &13 26 16.46 & -64 55 40.0  & - & 0.14610 & 0.11689 & - & 0.80007 & 15.49 - 15.63 I & \citet{2013AcA....63..379P}\\
OGLE-GD-DSCT-0054 & 13 31 35.98 & -64 09 35.7         & 0.10994 & 0.08539 &- & 0.77669 &- & 17.05 - 17.30 I & \citet{2013AcA....63..379P}\\
OGLE-GD-DSCT-0056 & 13 33 45.66 & -64 01 17.1         & - & 0.44410  & 0.35882 & - & 0.80797 & 15.63 - 15.73 I &\citet{2013AcA....63..379P} \\
LINEAR 9328902 & 13 35 49.76 & +26 55 16.7            & 0.05175 & 0.04047 &- & 0.78203 &- & 16.24 (0.45) V & \citet{2014PASP..126..509P}\\
GSC 03851-00240 & 13 45 21.66 & +54 11 51.2           & 0.06794 & 0.05260 &- & 0.77418 &- & 12.7 - 13.09 V &\citet{2015IBVS.6150....1W}\\
V1393 Cen & 13 57 15.60 & -52 55 22.6                 & 0.11778 & 0.09083 &- & 0.77118 &- & 9.16 - 9.63 V &   \citet{2009PZP.....9...26K} \\
GSC 02008-00003 & 14 22 31.21 & +24 34 57.0           & 0.05960 & 0.04614 &- & 0.77415 &- & 14.05 - 14.30 V & \citet{2010AA...520L..10B}\\
LINEAR 13836407 & 15 12 53.99 & +23 17 48.3           & 0.06001 & 0.04674 &- & 0.77884 &- & 16.78 - 17.23 V & \citet{2013AJ....146..101P} \\
V0638 Ser & 15 13 22.01 & +18 15 58.3                 & 0.05419 & 0.04191 &- & 0.77338 &- & 13.42 - 13.72 V & \citet{2014IBVS.6122....1W}  \\
CSS\_J151435.5-145958 & 15 14 35.52 & -14 59 58.4     & 0.04932 & 0.03845 & - & 0.77962 &- & 14.90 - 15.45 V & \citet{2009ApJ...696..870D} \\
QS Dra & 15 21 34.64 & +61 29 22.7                    & 0.09442 & 0.07304 &- & 0.77358 &- & 12.63 - 13.16 V & \citet{2007PZP....7....25K}\\
ASAS J152315-5603.7 & 15 23 15.43 & -56 03 43.2       & 0.12675 & 0.09767 &- & 0.77061 &- & 11.22 - 11.57 V & \citet{2014PZP....14....1K} \\
LINEAR 15289666 & 15 45 18.61 & +16 40 51.6           & 0.08654 & 0.06870 &- & 0.79383 &- & 15.98 - 16.42 V & \citet{2018PZ.....38....1K} \\
V0647 Ser & 15 52 51.39 & +06 06 06.1                 & 0.05492 & 0.04267 &- & 0.77689 & - &15.8 - 16.4 V  & \citet{2013AJ....146..101P}\\
GSC 2583-00504 & 16 13 31.71 & +32 34 42.8 &          0.05172 & 0.03999 &- & 0.77320 &- & 12.30 - 12.50 V &   \citet{2003IBVS.5442....1W} \\
LINEAR 16586778 & 16 13 57.55 & +28 28 57.2           & 0.07075 & 0.05570 &- & 0.78728 &- & 15.30 - 15.97 V & \citet{2009ApJ...696..870D} \\
V1553 Sco & 16 20 21.77 & -35 41 16.0                 & - & 0.18430  & 0.14704 & - & 0.79783 &  9.41 - 9.68 V & \citet{2009PZP.....9...26K}\\
CSS\_J162243.6+000503 & 16 22 43.63 & +00 05 03.1     & 0.10516 & 0.08130 &- & 0.77312 &- & 14.73 - 15.31 V & \citet{2018PZ.....38....1K} \\
CSS\_J162818.8+032651 & 16 28 18.88 & +03 26 51.1     & 0.12535 & 0.09859 &- & 0.78649 &- & 16.52 - 17.51 V & \citet{2018PZ.....38....1K} \\
V0552 Dra & 16 29 40.31 & +57 20 33.3                 & 0.06114 & 0.04750 &- & 0.77693 &- & 13.24 - 13.48 V & \citet{2014PZP....14....1K}\\
NSV 7805 & 16 32 20.12 & -02 12 08.3                  & 0.06460 & 0.05070 &- & 0.78477 &- & 16.2 - 16.8 V & \citet{2014ApJS..213....9D}\\
GSC 03887-00087 & 17 08 14.77 & +52 53 53.4           & 0.10718 & 0.08293 &- & 0.77374 &- & 13.45 - 13.80 V &  \citet{2010AA...520L..10B} \\
V879 Her & 17 31 12.72 & +28 03 16.8                  & 0.05689 & 0.04413 &- & 0.77564 &- & 15.23 - 15.88 V & \citet{2014IBVS.6122....1W} \\
V703 Sco & 17 42 16.81 & -32 31 23.6                  & 0.14996 & 0.11522 &- & 0.76832 &- & 7.58 - 8.04 V &   \citet{1966BAN....18..140O} \\
MGAB-V1182 & 17 45 32.66 & +28 45 51.7      & 0.05545 & 0.04333 &- & 0.78142 &- & 17.45 - 17.95 g & this work\\
CSS\_J174643.7+285533 & 17 46 43.81 & +28 55 33.3     & 0.08153 & 0.06346 &- & 0.77835 &- & 14.60 - 15.17 V & \citet{2018PZ.....38....1K}\\
NSV 9856 & 17 56 00.20 & -30 42 46.6                  & 0.11849 & 0.09127 & 0.07324 & 0.77028 & 0.80245 & 12.10 - 12.50 V & \citet{2014PZP....14....1K} \\
MACHO 119.19574.1169 & 18 02 00.37 & -29 48 43.2      & 0.10685 & 0.08272 &- & 0.77421 &- & 16.60 - 16.88 I & \citet{2000ApJ...536..798A}\\
OGLE BW1 V207 & 18 02 14.98 & -29 54 08.8             & 0.08560 & 0.06623 &- & 0.77375 &- & 18.67 V &  \citet{2005AA...440.1097P}\\
OGLE BW2 V142 & 18 02 18.04 & -30 08 11.4             & 0.06604 & 0.05140 &- & 0.77836 &- & 17.77 V &  \citet{2005AA...440.1097P}\\
MACHO 114.19840.890 & 18 02 31.85 & -29 27 03.9       & 0.12557 & 0.09679 &- & 0.77082 &- & 17.99 V & \citet{2000ApJ...536..798A}  \\
MACHO 114.19969.980 & 18 02 52.20 & -29 30 24.5       & 0.10327 & 0.07981 &- & 0.77282 &- & 17.03 - 17.35 I & \citet{2000ApJ...536..798A} \\
MACHO 128.21542.753 & 18 06 35.93 & -28 39 31.3       & 0.12005 & 0.09254 &- & 0.77083 &- & 17.79 V  & \citet{2000ApJ...536..798A} \\
MACHO 115.22573.263 & 18 09 00.48 & -29 14 30.9       & 0.09175 & 0.07087 &- & 0.77243 &- & 17.95 V & \citet{2000ApJ...536..798A} \\
MACHO 116.24384.481 & 18 13 16.45 & -29 49 27.0       & 0.08691 & 0.06716 &- & 0.77272 &- & 17.29 V &  \citet{2000ApJ...536..798A} \\
MACHO 162.25343.874 & 18 15 16.33 & -26 35 40.2       & 0.11128 & 0.08591 &- & 0.77196 &- & 17.96  V & \citet{2000ApJ...536..798A} \\
ATO J274.5653+06.5568 & 18 18 15.68 & +06 33 24.8     & 0.11151 & 0.08609 &- & 0.77204 &- & 13.80 - 14.35 V & \citet{2017PASP..129j4502K} \\
ASAS J182536-4213.6 & 18 25 36.26 & -42 13 35.8       & 0.10719 & 0.08216 &- & 0.76648 &- & 11.43 - 12.0 V & \citet{2011PZP....11....30K} \\
CzeV293 & 18 28 54.89 & +12 21 24.4                   & 0.05572 & 0.04312 &- & 0.77387 &- & 15.67 V & \citet{2016NewA...46...85S}\\
V575 Lyr & 18 29 43.24 & +28 09 54.6                  & 0.14556 & 0.11150 &- & 0.76602 &- & 12.55 - 12.85 V & \citet{2001IBVS.5205....1V} \\
V0836 Lyr$^{(c)}$ & 18 29 47.55 & +37 45 01.5 & 0.11658 & 0.09030 & 0.05084  & 0.77458 & 0.56308 & 12.7 - 13.0 V & \citet{2010AA...520L..10B} \\
MGAB-V1190 & 18 44 23.90 & +40 03 17.5      & 0.07205 & 0.05598 & - & 0.77696 &- & 17.40 - 17.60 g & 
this work\\
SSS\_J184425.3-563648 & 18 44 25.40 & -56 36 46.2     & 0.09437 & 0.07283 &- & 0.77169 &- & 13.58 - 13.99 V & \citet{2018PZ.....38....1K} \\
KIC 5950759 & 19 15 00.54 & +41 13 55.4               & 0.07032 & 0.05453   &- & 0.77555 & - &13.516 V &  \citet{2018ApJ...863..195Y}\\
ASAS J192227-5622.5 & 19 22 27.39 & -56 22 28.1       & 0.14909 & 0.11277 &- & 0.75639 &- & 11.64 - 12.08 V & \citet{2011PZP....11....30K} \\
KIC 10975348 & 19 26 46.11 & +48 25 30.8              & 0.09773 & 0.07408 &- & 0.75799 &- & 18.598 (0.7) Kp & \citet{2021AJ....161...27Y}\\
KIC 2857323 & 19 29 49.16 & +38 01 21.7               & 0.07618 & 0.05897 &- & 0.77409 &- & 13.342 (0.27) Kp&  \citet{2018AA...614A..46B}\\
V0798 Cyg & 19 38 06.90 & +30 54 33.5                 & - & 0.19477  & 0.15602 & - & 0.80105 & 12.28 - 12.66 V & \citet{1998PASP..110.1156M}\\
SSS\_J194149.5-533246 & 19 41 49.54 & -53 32 46.6     & 0.11106 & 0.08578 &- & 0.77236 &- & 15.70 - 16.14 V & \citet{2018PZ.....38....1K} \\
KIC 10284901 & 19 43 46.28 & +47 20 32.8              & 0.05265 & 0.04109 &- & 0.78044 &- & 15.459 Kp & \citet{2019ApJ...879...59Y}  \\
V829 Aql & 19 46 57.29 & +03 30 28.5                  & 0.29244 & 0.22097 & 0.17650 & 0.75561 & 0.79875 & 10.00 - 10.50 V & \citet{1998IBVS.4549....1H} \\
GSC 03144-595 & 19 48 02.92 & +41 46 55.8             & 0.20364 & 0.15549 & 0.12445 & 0.76355 & 0.80037 & 10.45 - 10.95 V & \citet{2016AJ....152...17M}\\
GSC 06905-01641 & 20 10 22.51 & -23 10 59.7           & 0.06880 & 0.05329 &- & 0.77459 &- & 14.84 - 15.29 V & \citet{2014PZP....14....1K} \\
GSC 03949-00386 & 20 19 44.95 & +58 29 20.0           & 0.09578 & 0.07394 &- & 0.77193 &- & 11.0 - 11.2 V & \citet{2012IBVS.6013....1B} \\
GSC 03949-00811 & 20 26 01.74 & +59 30 53.5           & 0.16975 & 0.13008 &- & 0.76629 &- & 11.2 V & \citet{2005AA...440.1097P} \\
GSC 07460-01520 & 20 33 38.54 & -32 55 03.6           & 0.08701 & 0.06815 &- & 0.78326 &- & 14.30 - 14.65 V & \citet{2010AA...520L..10B} \\
CSS\_J205540.0-020320 & 20 55 40.03 & -02 03 21.2     & 0.08394 & 0.06556 &- & 0.78097 &- & 15.67 - 16.40 V & \citet{2018PZ.....38....1K} \\
VY Equ & 20 58 49.64 & +08 54 05.3                    & - & 0.17645  & 0.14078 & - & 0.79785 & 10.21 - 10.51 V & \citet{2011PZP....11....30K}\\
GSC 05183-01938 & 20 59 10.15 & -03 28 20.7           & 0.07622 & 0.05902 &- & 0.77437 &- & 14.45 - 15.11 V & \citet{2018PZ.....38....1K} \\
\lbrack SIG2010\rbrack~3269918 &20 59 27.28 &-01 13 49.0 & 0.05238 & 0.04089 &- & 0.78061 &- & 17.68 (0.49) r& \citet{2012MNRAS.424.2528S}\\
V1719 Cyg & 21 04 32.92 & +50 47 03.3                 & - & 0.26730  & 0.21378 & - & 0.79978 &  7.81 -  8.21 V & \citet{1986AA...158..389M}\\
V3124 Cyg & 21 12 53.69 & +33 17 34.3                 & - & 0.10530  & 0.08785 & - & 0.83428 & 10.47 - 10.69 V & \citet{2011PZP....11....30K} \\
GSC 04257-00471 & 21 26 01.11 & +64 30 57.5           & 0.17380 & 0.13308 &- & 0.76574 &- & 11.4 V & \citet{2005AA...440.1097P} \\
CSS\_J212609.1+145319 & 21 26 09.20 & +14 53 18.7     & 0.09042 & 0.07054 &- & 0.78017 &- & 13.99 - 14.74 V & \citet{2018PZ.....38....1K} \\
V0371 Aqr & 21 29 52.69 & -01 10 18.9                 & 0.08059 & 0.06244 &- & 0.77479 &- & 14.22 - 14.65 V & \citet{2013PZP....13...11K} \\
BP Peg & 21 33 13.53 & +22 44 24.3                    & 0.10954 & 0.08451 &- & 0.77148 &- & 11.69 - 12.28 V & \citet{1983GEOCR...6.....F,2014NewA...32....6W} \\
GSC 01128-00127 & 21 35 32.99 & +12 43 41.3           & 0.06305 & 0.04878 &- & 0.77359 &- & 14.73 - 15.12 V& \citet{2009ApJ...696..870D} \\
GSC2.3 SBA9013761 & 21 36 30.17 & -00 21 27.6         & 0.10740 & 0.08368 &- & 0.77907 &- & 18.19 (0.22) r & \citet{2012MNRAS.424.2528S} \\
CSS J213934.3-050020 & 21 39 34.33 &-05 00 20.3       & 0.14664 & 0.11273 &- & 0.76875 &- & 13.90 V & \citet{2021PASP..133e4201S}\\
CSS\_J214745.8+122726 & 21 47 45.78 & +12 27 26.6     & 0.07820 & 0.06062 &- & 0.77518 &- & 14.18 - 14.36 V & \citet{2009ApJ...696..870D} \\
DDE 147 & 21 53 05.55 & +23 53 03.2                   & 0.06340 & 0.04987 &- & 0.78656 &- & 14.15 - 14.55 V &\citet{2009ApJ...696..870D} \\
USNO-A2.0 1425-12623576 & 21 59 23.24 & +59 24 56.9   & 0.10273 & 0.07917 &- & 0.77061 &- & 14.33 - 14.72 R & \citet{2015PZP....15....8L}\\
AG Aqr & 22 05 31.82 & -22 30 00.7                    & 0.29174 & 0.22220 &- & 0.76165 & - & 14.58 - 14.96 V & \citet{2013PZP....13...11K} \\
DDE 148 & 22 20 39.22 & +23 43 13.8                   & 0.11261 & 0.08699 &- & 0.77247 &- & 13.23 - 13.73 V &  \citet{2009ApJ...696..870D} \\
GSC 06385-01170 & 22 24 23.25 & -15 38 05.5           & 0.05718 & 0.04453 &- & 0.77881 &- & 14.371 (0.487) V& \citet{2009ApJ...696..870D}\\
DDE 145 & 22 25 27.04 & +08 00 13.8                   & 0.05451 & 0.04235 &- & 0.77690 &- & 14.20 - 14.70 V & \citet{2009ApJ...696..870D} \\
ASAS J231801-4520.0 & 23 18 01.14 & -45 19 55.0       & 0.11501 & 0.08892 &- & 0.77313 &- & 12.79 - 13.74 V & \citet{2015PZP....15....9K}\\
V0761 Peg & 23 20 56.45 & +34 51 50.9                 & 0.14194 & 0.10925 & 0.08767 & 0.76969 & 0.80247 & 13.26 - 13.50 R & \citet{2002ApJ...577..845K} \\
CSS\_J235659.3+063132 & 23 56 59.35 & +06 31 32.5     & 0.09470 & 0.07327 &- & 0.77371 &- & 14.91 - 15.39 V & \citet{2018PZ.....38....1K}\\

\enddata

Note: A list of 155 double- and multi-mode HADS, including 132 stars pulsating in both P0 and P1, ten stars in both P1 and P2, 11 triple-mode HADS, and 2 quadruple-mode HADS. (a) this star is a quadruple-mode HADS with P0, P1, P2 and P3(= 0.08960 days). (b) this star is also a triple-mode pulsator, but with P1, P2, and P3(=0.09742 days). (c) this star is a quadruple-mode HADS with P0, P1, P2 and P3(= 0.05829 days).

\end{deluxetable}
\end{landscape}

\section{Statistical analysis of the sample}

\subsection{The distribution of the fundamental period (P0) and the period ratio P1/P0 of HADS(B)}

\begin{figure*}
\begin{center}
  \includegraphics[width=0.9\textwidth, trim = 20 245 30 240, clip]{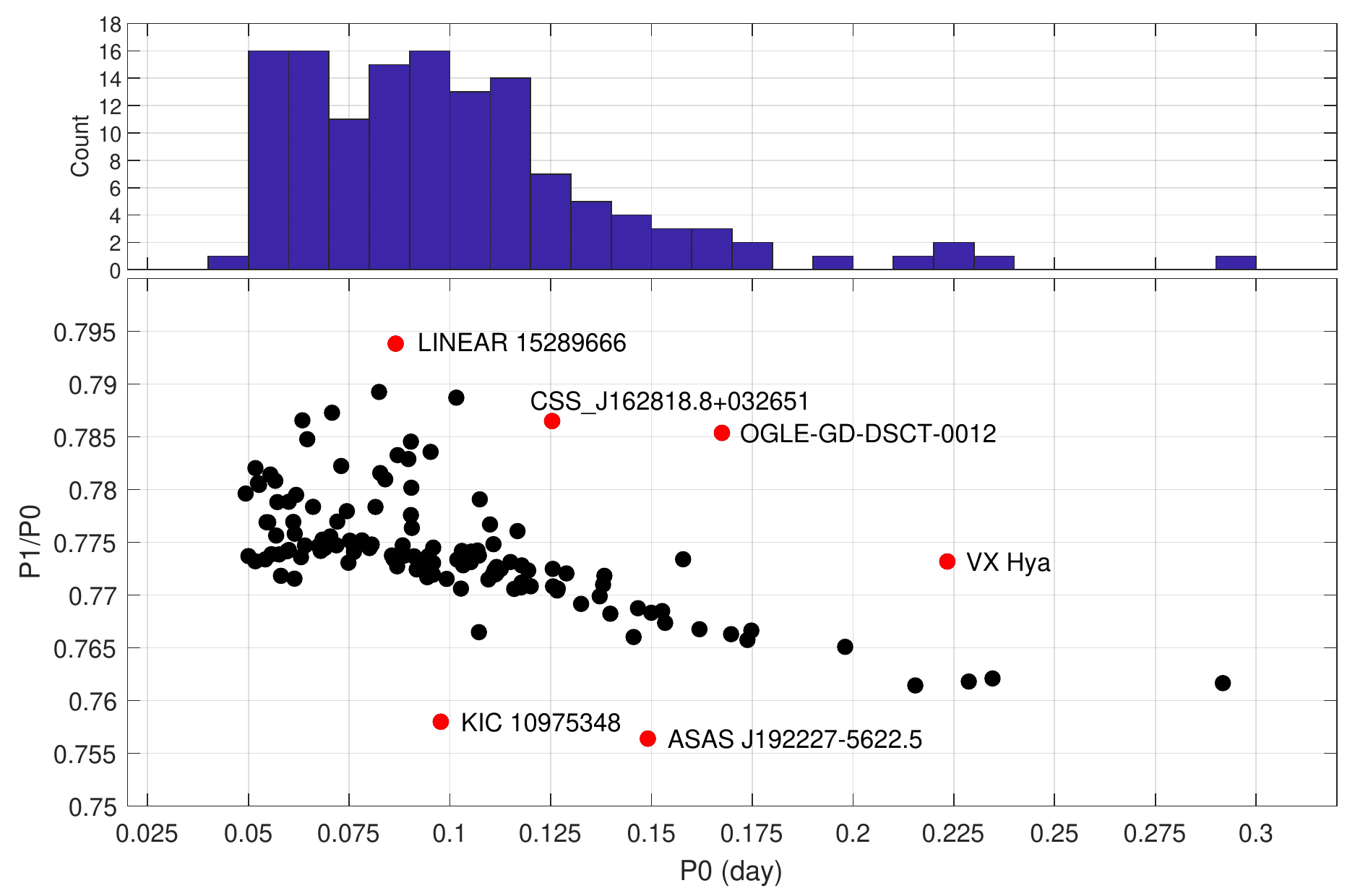}
  \caption{Top panel: the distribution of P0 of 132 stars pulsating in F and 1O. It is clear that most of the stars lie in a range of 0.05 days to 0.175 days. Bottom panel: Petersen diagram of P1/P0 and P0. The red dots indicate six stars with unusual period ratios.}
    \label{fig:data_points}
\end{center}
\end{figure*}

In Figure \ref{fig:data_points}, we plot a histogram of the fundamental period (P0) with 132 stars pulsating in F+1O, as well as a Petersen diagram of P0 and the ratio P1/P0. 
We found most of the stars are clearly grouped in a range of 0.05 days to 0.175 days. In the period range of 0.12 days to 0.175 days, the number of stars shows a clear decreasing trend, and if P0 > 0.18 days, there are only six stars known at present.

Moreover, the distribution of stars shows a bimodal structure, which is not very significant but with possible peaks at P0 $\sim$ 0.06 days and 0.09 days. The underlying reason is unclear, but possibly due to the currently limited sample. We suggest more such sample particularly in short periods is still needed to further verify this feature.

From Petersen diagram in Figure \ref{fig:data_points}, we note six stars (i.e. LINEAR 15289666, KIC 10975348, CSS\_J162818.8+032651, ASAS J192227-5622.5, OGLE-GD-DSCT-0012, VX Hya) deviate largely from the downward trend of P1/P0, compared with the corresponding P0. To help readers investigate these stars easily, a brief introduction to these six stars is listed below:

$\bullet$ LINEAR 15289666: Its value of P1/P0 is 0.7938, which is higher than that of all others. The variability was discovered by Palaversa et al. (2013). Drake et al. (2014) detected the star's variability independently and gave the type HADS with period of 0.86539 days. This star is actually a HADS(B), but it is difficult to determine the correct periods due to the one-day alias. The currently determined period ratio (P1/P0 = 0.7938) is clearly higher than typical value of P1/P0. \citet{2018PZ.....38....1K} noticed that if considering the one-day-alias corrected periods (P0$_{c}$=0.094763 days, P1$_{c}$=0.0737655 days), the corrected period ratio, P1$_{c}$/P0$_{c}$ = 0.7784, is more typical. Continuous photometric observations from space missions and/or multi-site campaigns could provide more reliable periods.

$\bullet$ KIC 10975348: Its value of P1/P0 is 0.7580, which is a little lower than that of other stars with the same P0. It was classified as a $\delta$ Sct star with a pulsation period of 0.0979 days by \citet{2014MNRAS.437..132R}. Subsequently, \citet{2021AJ....161...27Y} analyzed the pulsating behavior of this star using the time-series data from Kepler mission and reported two new independent frequencies F1(=13.4988 c/day) and F2(=19.0002 c/day). It has a relatively low period ratio of 0.7580, revealing that it might be a metal-rich variable star. The frequency F2 may be a third overtone, making this star a radial triple-mode HADS candidate. The $O-C$ analysis suggests this star seem to show no obvious period change, which is in contrast to the majority of HADS. To understand why it has a lower period ratio and verify the period variations, seismic modelling and regular observations from space with a longer time span are needed in the future.

$\bullet$ CSS\_J162818.8+032651: the value of P1/P0 is 0.78649, which is higher than that of the stars with similar P0. It was classified as a HADS with a fundamental period of 0.125352 days by \citet{2014ApJS..213....9D}. Subsequently, \citet{2018PZ.....38....1K} re-analyzed the light variations of this star using the photometric data from the Catalina Sky Surveys \citep[CSS,][]{2009ApJ...696..870D} and identified it as a double-mode HADS. There is no more available information about this star at present. To explore the reason for its relatively higher period ratio, seismic modelling and spectroscopic observations  are needed in the future.

$\bullet$ ASAS J192227-5622.5: the value of P1/P0 is 0.7639, which is the lowest value in the present sample. The variability of this star was reported by \citet{2002AcA....52..397P}. Subsequently, \citet{2011PZP....11....30K} re-analyzed this star using ASAS-3 data \citep{2002AcA....52..397P} and classified it as a double-mode HADS. There is no more available information about this star at present. To explore the reason for its lowest period ratio, seismic modelling and spectroscopic observations are needed in the future.

$\bullet$ OGLE-GD-DSCT-0012: the value of P1/P0 is 0.78537, which is higher than that of the stars with similar P0. This star was discovered by \citet{2013AcA....63..379P} in OGLE-III Galactic disk area. \citet{2013AcA....63..379P} found the period ratio of this star cannot be reproduced by their theoretical models. Subsequently, \citet{2015AcA....65...63P} presented a spectroscopic follow-up observation for this star and gave the spectral type as F0. There is no more photometry available for this star at present. To explore the reason for its higher period ratio, seismic modelling and spectroscopic observations are needed  in the future.

$\bullet$ VX Hya: it was discovered by \citet{1931AN....242..129H} and then observational studies were performed by many researchers \citep{1932AN....244..417L,1938BAN.....8..277O,1966ApJ...143..852F,1977PASP...89...55B}. \citet{2009PASP..121.1076T} determined its frequency of the fundamental mode as $f_{0}$ = 4.4765 c/day and the first overtone $f_{1}$ = 5.7898 c/day through a two-year, multi-site observing campaign. Recently, \citet{2018ApJ...861...96X} derived its mass as 2.385 $\pm$ 0.025 M$_{\odot}$ with asteroseismic models and found it is located at the post-MS stage with a helium core and a hydrogen-burning shell in H-R diagram. Its higher mass and late evolutionary stage may be responsible for its higher ratio of 0.77318, compared with other five stars.

\begin{figure*}
\begin{center}
  \includegraphics[width=0.9\textwidth, trim = 40 245 56 265, clip]{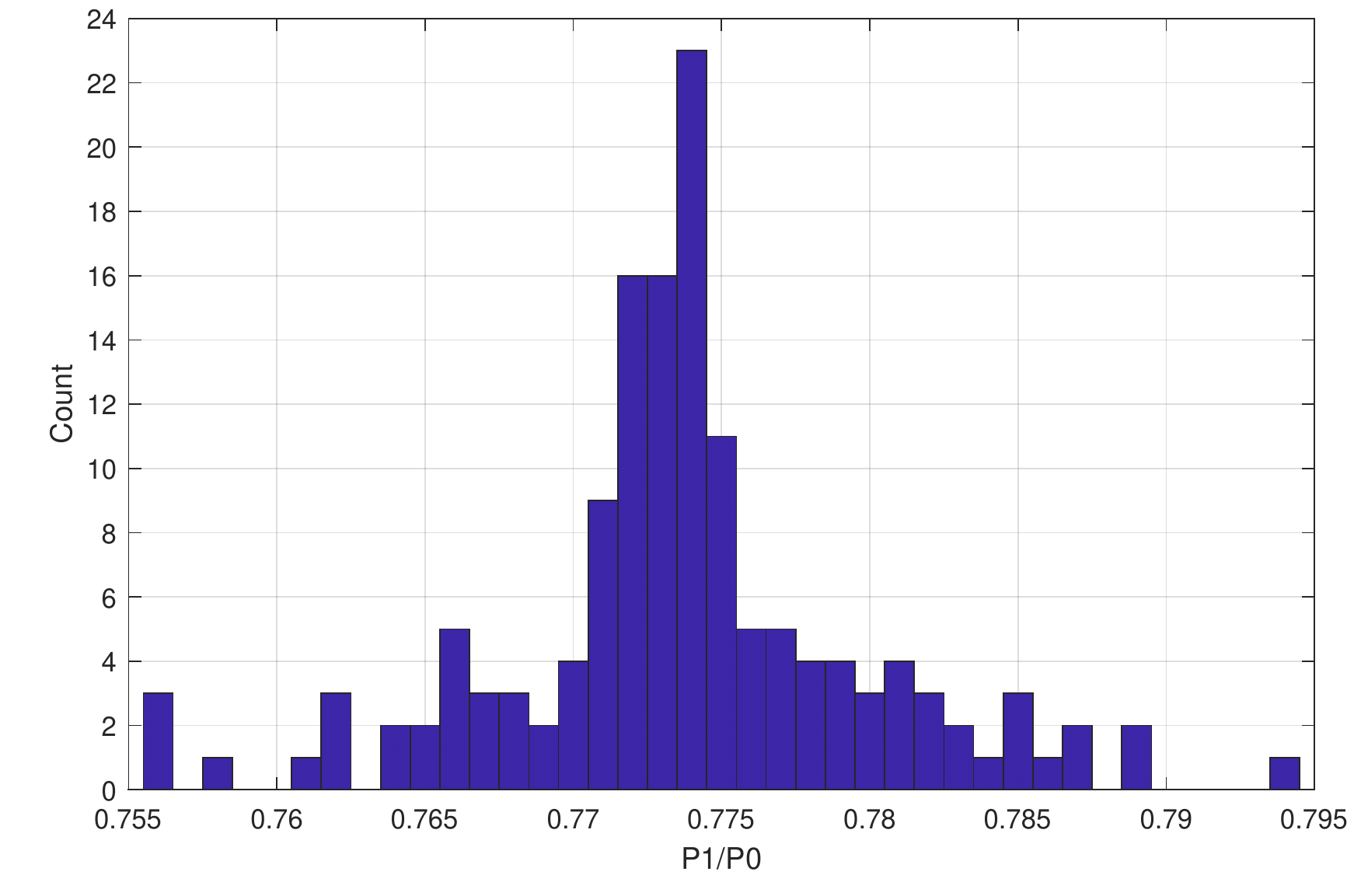}
  \caption{The distribution of period ratio P1/P0 of 144 stars with both P0 and P1. It clearly shows that most stars lie in a wide range of 0.761 - 0.787, compared to that in \citet{2000ASPC..210....3B}.}
    \label{fig:P1_P0_distribution}
\end{center}
\end{figure*}

The period ratio P1/P0 has long been taken as the critical parameter to identify a HADS(B). The typical value of P1/P0 is about 0.77, which might be sightly shifted depending on metallicity and/or mass of the star \citep{2005AA...440.1097P}. Using a much larger sample (142 HADS(B) with both P0 and P1), we further investigated the distribution of P1/P0 and created a histogram of P1/P0, \textbf{as shown in Figure \ref{fig:P1_P0_distribution}}. We found that most stars lie in a range of 0.761 - 0.787, in which 87.8 percent of the stars concentrate in a range of 0.766 - 0.785, which is slightly wider than the previous studies (0.77 - 0.78 of \citealt{2000ASPC..210....3B}). And in this narrow range, the ratio P1/P0 seems to have a peak at P1/P0 = 0.774, which is slightly higher than the value of 0.772 in the metal-rich HADS stars presented by \citet{2000ASPC..210..373M}. We note the stars with the highest and lowest values of P1/P0 have been listed above and may need more attention to their pulsations.

\subsection{The relationship between P1/P0 and P0 of the HADS(B) stars}

Petersen Diagram, showing the period ratio of 1O/F as a function of the fundamental mode, is a well-known tool to study radial pulsators, especially double-mode pulsators \citep{1973AA....27...89P,1978AA....62..205P}. For HADS(B), the relationship between the fundamental radial mode and period ratio of 1O/F was first studied by \cite{1996A&A...312..463P} using 7 HADS(B). In their work, a series of theoretical models were created and the model predicted a peak value of P1/P0 around 0.774 when log P0 equals $-$0.9. For both increasing and decreasing value of P0, the value of P1/P0 shows a downward trend, i.e. P1/P0 equals 0.764 for the value of log P0 is about $-$0.55, and equals 0.770 for log P0 = $-$1.1. In the case of extreme simplification of the model, the relation could be treated as a downward parabola shape with vertex 0.774 with log P0 = $-$0.9 for the HADS(B) \citep{1996A&A...312..463P}.

\begin{figure*}
\begin{center}
  \includegraphics[width=0.9\textwidth, trim = 40 250 40 265, clip]{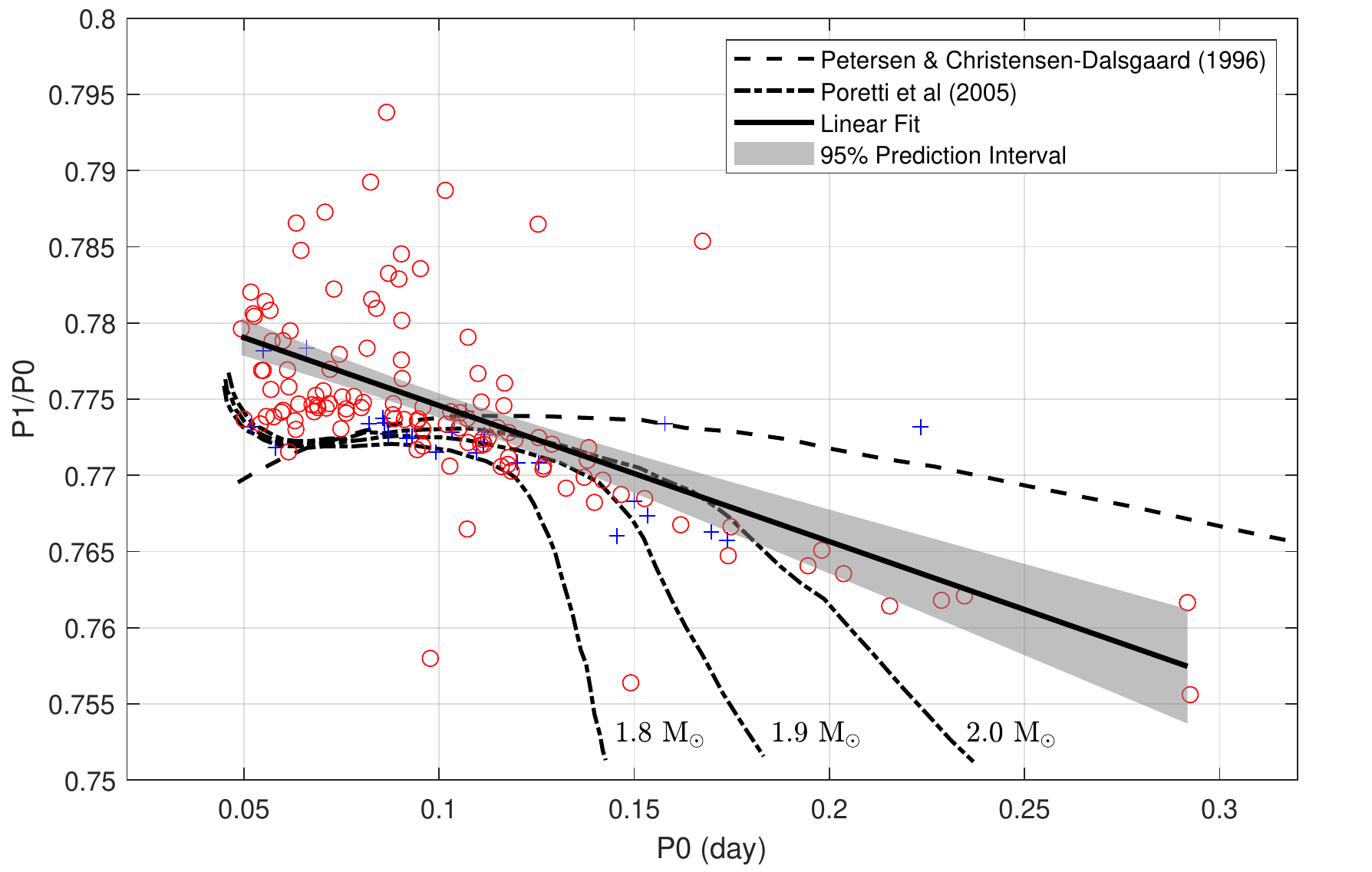}
  \caption{Petersen diagram of 144 HADS(B) stars with both P0 and P1. The plus signs indicate 25 stars used in \cite{2005AA...440.1097P}. Three dash-dotted lines are the theoretical sequences calculated by \cite{2005AA...440.1097P}. The dashed line is the relationship given by \cite{1996A&A...312..463P} using seven stars. The solid line with shaded area is the best-fitted line in this work.}
    \label{fig:new_P1_P0_P0}
\end{center}
\end{figure*}

Subsequently, with 25 HADS(B), \cite{2005AA...440.1097P} calculated a new set of theoretical models and found the period ratio P1/P0 decreases toward longer periods overall. Specifically, the decreasing trends in the theoretical sequences remain a long standstill between P0 = 0.05 days and P0 = 0.126 days, while for stars with longer period P0 > 0.126 days, the ratio decreases rapidly, as shown in Figure \ref{fig:new_P1_P0_P0}. During the period standstill, \citet{2005AA...440.1097P} found the period ratios are more sensitive to the changes in metallicity than in mass, but at longer periods, small difference in mass can strongly influence the period ratio.

Using a larger sample of 77 HADS(B), \citet{2016JAVSO..44....6F} pointed out there seems to exist a linear relation between the ratio P1/P0 and the period P0, and gave a relation as: Y = $-$ 0.084809($\pm$ 0.008298) P0 + 0.782048($\pm$ 0.000995), where Y is the period ratio P1/P0, and P0 is the radial fundamental period in days.

To further investigate the relationship between P1/P0 and P0 of HADS(B), we plot a Petersen diagram with 144 stars with both P0 and P1 (i.e. stars pulsating in F+1O, F+1O+2O, and F+1O+2O+3O, as listed in Table \ref{tab:List1}). Figure \ref{fig:new_P1_P0_P0} shows the Petersen diagram in this work. To explore the relation between P1/P0 and P0, we attempt to make a linear-fit for these stars, excluding six unusual stars mentioned above. The corresponding linear equation was derived as: 
\begin{equation} \label{eq:linefit}
Y = - 0.0940(\pm 0.0091) P0 + 0.7835( \pm 0.0009), 
\end{equation}
with a correlation coefficient of 0.6865. This possible linear relation is similar to the result in \citet{2016JAVSO..44....6F}. Although \citet{2016JAVSO..44....6F} believe that more HADS(B) could improve the parameters of the possible linear relation, the derived parameters in our study are only sightly higher than that obtained by \citet{2016JAVSO..44....6F}.

For the theoretical relationship between P0 and P1/P0 derived by \cite{1996A&A...312..463P}, we find that most of HADS(B) in our sample are not in agreement with the predicted values. As a comparison, we also plot the theoretical sequences calculated by \cite{2005AA...440.1097P} in Figure \ref{fig:new_P1_P0_P0}, as well as 25 HADS(B) in their work. From the figure, most of the stars are far away from the theoretical predicted values in periods of 0.05 to 0.10 days. Between periods of 0.10 and 0.20 days, most of the stars are essentially in agreement with the predicted values. But if the period is larger than 0.20 days, three stars (i.e., VX Hya, AG Aqr, and V829 Aql)are far away from the predictions.

For the scattering of P1/P0 with P0 between 0.05 and 0.10 days, one of the possible reasons may be stellar rotation. For instance, \citet{2006A&A...447..649S} investigated the effect of rotation on period ratios for double-mode pulsators and found that the differences in period ratios increase with the rotational velocities for a given metallicity. Moreover, it was also found that the differences in period ratios caused by rotation are equivalent to the differences by metallicity up to 0.30 dex, and it should be taken into account when one wants to accurately determine the mass and metallicity of a star. On the other hand, \citet{2007A&A...474..961S} also investigated the influence of near-degeneracy on period ratios and found that near-degeneracy could change the oscillation frequencies through the coupling strength and thereby modified the period ratios significantly even for the relatively slow rotators. In terms of frequency variations, the coupling strength analysis suggests that near-degeneracy may modify the frequency of the fundamental radial mode and the first overtone up to 0.3 $\mu$Hz.

Therefore, for the scatters far away from the typical values of P1/P0, especially in short periods between 0.05 and 0.10 days in Figure \ref{fig:new_P1_P0_P0}, the possible reasons for their abnormality may include the differences of stellar mass, rotational velocity, and metallicity, as well as other physical parameters. On the other hand, we cannot exclude the possibility that these scatters are due to a pair of consecutive radial overtones, radial and non-radial mode or even two non-radial modes \citep{2002ApJ...576..963T,2006MmSAI..77..223P}. We suggest detailed seismic modelling for these stars considering these factors is needed, which is helpful to understand their peculiarity and the internal physics in the future.

\subsection{The period ratios of multi-mode HADS}

In this catalogue, there are 13 HADS pulsating in multi-mode, i.e. 11 star with triple-mode and 2 with quadruple-mode (F+1O+2O+3O). In addition, there are also 10 double-mode HADS pulsating in 1O+2O. These stars provide a good opportunity to explore the period ratios between the radial overtones and the fundamental mode.

\begin{figure*}
\begin{center}
  \includegraphics[width=0.9\textwidth, trim = 40 245 50 265, clip]{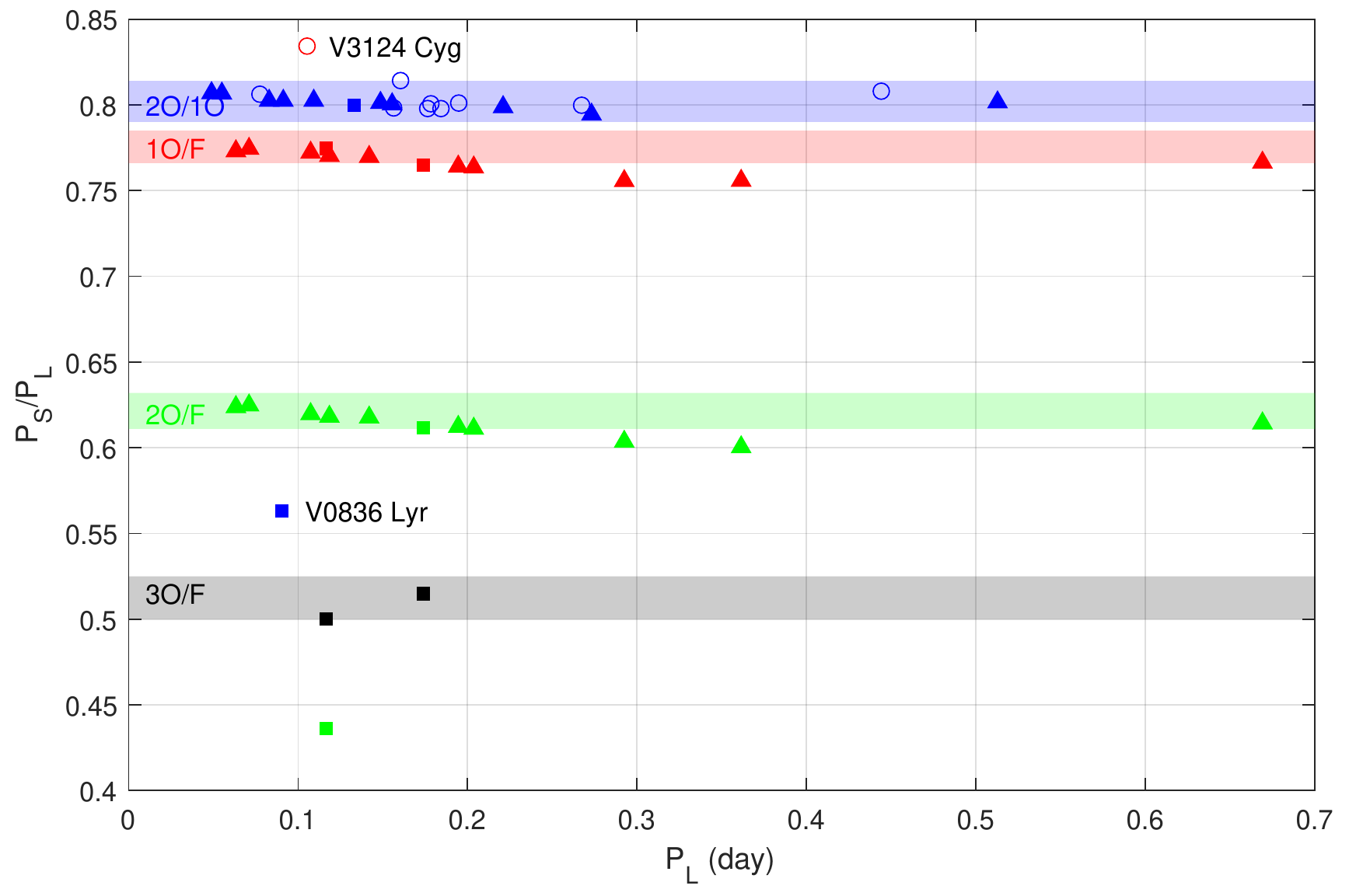}
  \caption{Petersen diagram for multi-mode pulsators in this work. Open circles, triangles and squares correspond to the double-mode HADS with P1 and P2, triple-mode and quadruple-mode HADS, respectively. The blue strip shows the ratios of 2O/1O, which lie in a range of 0.802 $\pm$ 3 $\sigma$, $\sigma$ is 0.004, derived in this work. The red strip shows the ratios of 1O/F determined in Sec. 3. The green and grey stand for the ratios of the radial second-overtone and third-overtone to the fundamental mode, respectively.}
    \label{fig:P2_P1_P1}
\end{center}
\end{figure*}

Figure \ref{fig:P2_P1_P1} shows a Petersen diagram for multi-mode HADS. From this figure, we found that almost all the ratios of P2/P1 lie in a narrow horizontal strip around the value of 0.80, except for two points in obviously puzzling positions from stars V3124 Cyg and V0836 Lyr. If these two outliers are excluded, we estimate the mean value of P2/P1 of the 21 stars is 0.802 with standard deviation of 0.004. In this strip, the distribution of P2/P1 do not show any up or down trend. The ratios of P1/P0 of these stars are also within the typical ratio range of 0.761 - 0.787, in consistent with that of HADS(B), suggesting there seems no difference in terms of P1/P0 between multi-mode HADS and usual HADS(B). For the values of P2/P0 and P3/P0, these stars are also nearly in the range of period ratios predicted by \citet{1979ApJ...227..935S} and \citet{2021arXiv210708064N}.

A brief introduction to the two stars showing unusual period ratio P2/P1 in Figur \ref{fig:P2_P1_P1} is presented below:

$\bullet$ V3124 Cyg: \citet{2014PZP....14....1K} provided two pulsating periods and their modes: 0.10530 days (P1) and 0.08785 days (P2) and the P2/P1 ratio is 0.834. We re-analyzed the available light curve and found the amplitude of P2 is larger than that of P1. The pulsations might be caused by the second-overtone and third-overtone modes. Further observations and seismic modellings are needed to verify the modes.

$\bullet$ V0836 Lyr: four independent frequencies are listed in the database, but other periods with lower amplitudes still may exist. Its P1/P0 ratio is 0.77458, but P2/P1 equals 0.56308, which is inconsistent with the typical value (i.e., 0.80) of P2/P1. \textbf{Also, the period ratio of 2O/F (0.05084/0.11658=0.4361) is far away from the typical value of 2O/F.} So we suspect the P2 is not correctly identified at present. Further analysis with seismic modellings may help to identify the modes.

\section{Summary}

In this work, we collected all the double-mode and multi-mode HADS as far as possible, and first created a catalogue with a sample of 155 stars, including 142 double-mode HADS (i.e. 132 stars pulsating in both the fundamental radial mode and first-overtone, 10 stars in both the first-overtone and second-overtone), 11 triple-mode HADS, and 2 quadruple-mode HADS.

Statistical analysis of the 132 double-mode HADS pulsating in both P0 and P1 indicate that most of them are pulsating with a fundamental period (P0) in the range of 0.05 days to 0.175 days, and the distribution of P0 shows a hint of bimodal structure with peaks at 0.06 days and 0.09 days, which still needs further investigation. Using 142 stars with both P0 and P1, we investigated the distribution of P1/P0 and find that the values of P1/P0 lie in a wide range of 0.761 $-$ 0.787. Nearly 90 percent of the stars concentrate in a relatively narrow range of 0.765 $-$ 0.785.

With 144 HADS(B) collected in this work, we created a new Petersen diagram between P1/P0 and P0, and derived an updated linear relation between P1/P0 and P0 as: P1/P0 = $-$0.0940($\pm$ 0.0091) P0 + 0.7835( $\pm$ 0.0009), with a correlation coefficient of 0.6865. Compared to the result obtained by \citet{2016JAVSO..44....6F}, we found the new result is essentially in agreement with theirs, only with slightly higher values in both the slope and intercept of the linear relation.

We also examined the relation between P2/P1 and P1 for 23 HADS pulsating in both P1 and P2, and found the values of P2/P1 fall in a horizontal strip with mean value of 0.802 and standard deviation of 0.004, if the two obvious outliers were excluded. It seems to indicate that this narrow strip could be considered as a possible observational indicator to identify the modes 1O and 2O for the multi-mode HADS.

\acknowledgements

This research is supported by the National Natural Science Foundation of China (grant Nos. 11573021, U1938104, 12003020) and the Fundamental Research Funds for the Central Universities. We would like to thank the $AAVSO$ $International$ $Variable$ $Star$ $Index$ for providing such data.


\begin{thebibliography}{}

\bibitem[Aerts et al.(2010)]{2010aste.book.....A} Aerts, C., Christensen-Dalsgaard, J., \& Kurtz, D.~W.\ 2010, Asteroseismology, Astronomy and Astrophysics Library. ISBN 978-1-4020-5178-4. Springer Science+Business Media B.V., 2010, p.

\bibitem[Alcock et al.(1997)]{1997ApJ...486..697A} Alcock, C., Allsman, R.~A., Alves, D., et al.\ 1997, \apj, 486, 697.

\bibitem[Alcock et al.(2000)]{2000ApJ...536..798A} Alcock, C., Allsman, R.~A., Alves, D.~R., et al.\ 2000, \apj, 536, 798.

\bibitem[Allsman \& Axelrod(2001)]{2001astro.ph..8444A} Allsman, R.~A. \& Axelrod, T.~S.\ 2001, astro-ph/0108444

\bibitem[Antoci et al.(2014)]{2014ApJ...796..118A} Antoci, V., Cunha, M., Houdek, G., et al.\ 2014, \apj, 796, 118.

\bibitem[Antoci et al.(2019)]{2019MNRAS.490.4040A} Antoci, V., Cunha, M.~S., Bowman, D.~M., et al.\ 2019, \mnras, 490, 4040


\bibitem[Bates \& Halliwell(1985)]{1985AA...151..403B} Bates, B. \& Halliwell, D.~R.\ 1985, \aap, 151, 403

\bibitem[Balona \& Nemec(2012)]{2012MNRAS.426.2413B} Balona, L.~A. \& Nemec, J.~M.\ 2012, \mnras, 426, 2413.

\bibitem[Balona(2016)]{2016MNRAS.459.1097B} Balona, L.~A.\ 2016, \mnras, 459, 1097. doi:10.1093/mnras/stw671

\bibitem[Barcel{\'o} Forteza et al.(2018)]{2018AA...614A..46B} Barcel{\'o} Forteza, S., Roca Cort{\'e}s, T., \& Garc{\'\i}a, R.~A.\ 2018, \aap, 614, A46. 

\bibitem[Bellm et al.(2019)]{2019PASP..131a8002B} Bellm, E.~C., Kulkarni, S.~R., Graham, M.~J., et al.\ 2019, \pasp, 131, 018002.

\bibitem[Bernhard et al.(2004)]{2004IBVS.5552....1B} Bernhard, K., Pejcha, O., Proksch, W., et al.\ 2004, Information Bulletin on Variable Stars, 5552, 1

\bibitem[Bernhard et al.(2012)]{2012IBVS.6013....1B} Bernhard, K., Srdoc, G., \& Frank, P.\ 2012, Information Bulletin on Variable Stars, 6013, 1

\bibitem[Borucki et al.(2010)]{2010Sci...327..977B} Borucki, W.~J., Koch, D., Basri, G., et al.\ 2010, Science, 327, 977.


\bibitem[Bowman(2017)]{2017ampm.book.....B} Bowman, D.~M.\ 2017, Amplitude Modulation of Pulsation Modes in Delta Scuti Stars, Springer Theses series. ISBN 978-3-319-66649-5. Springer International Publishing, 2017. 

\bibitem[Bowman et al.(2021)]{2021MNRAS.504.4039B} Bowman, D.~M., Hermans, J., Daszy{\'n}ska-Daszkiewicz, J., et al.\ 2021, \mnras, 504, 4039.

\bibitem[Breger(1975)]{1975ApJ...201..653B} Breger, M.\ 1975, \apj, 201, 653.

\bibitem[Breger(1977)]{1977PASP...89...55B} Breger, M.\ 1977, \pasp, 89, 55.

\bibitem[Breger(1980)]{1980ApJ...235..153B} Breger, M.\ 1980, \apj, 235, 153.

\bibitem[Breger(2000)]{2000ASPC..210....3B} Breger, M.\ 2000, Delta Scuti and Related Stars, 210, 3

\bibitem[Broglia(1959)]{1959MmSAI..30...57B} Broglia, P.\ 1959, \memsai, 30, 57

\bibitem[Butters et al.(2010)]{2010AA...520L..10B} Butters, O.~W., West, R.~G., Anderson, D.~R., et al.\ 2010, \aap, 520, L10.

\bibitem[Chang et al.(2013)]{2013AJ....145..132C} Chang, S.-W., Protopapas, P., Kim, D.-W., et al.\ 2013, \aj, 145, 132.

\bibitem[Coates et al.(1980)]{1980IBVS.1756....1C} Coates, D.~W., Halprin, L., Heintze, G.~N., et al.\ 1980, Information Bulletin on Variable Stars, 1756, 1

\bibitem[Drake et al.(2009)]{2009ApJ...696..870D} Drake, A.~J., Djorgovski, S.~G., Mahabal, A., et al.\ 2009, \apj, 696, 870.

\bibitem[Drake et al.(2014)]{2014ApJS..213....9D} Drake, A.~J., Graham, M.~J., Djorgovski, S.~G., et al.\ 2014, \apjs, 213, 9.

\bibitem[Drake et al.(2017)]{2017MNRAS.469.3688D} Drake, A.~J., Djorgovski, S.~G., Catelan, M., et al.\ 2017, \mnras, 469, 3688.

\bibitem[Eggen(1970)]{1970PASP...82..274E} Eggen, O.~J.\ 1970, \pasp, 82, 274.

\bibitem[Eggen(1979)]{1979ApJS...41..413E} Eggen, O.~J.\ 1979, \apjs, 41, 413.

\bibitem[Eggen \& Iben(1989)]{1989AJ.....97..431E} Eggen, O.~J. \& Iben, I.\ 1989, \aj, 97, 431.

\bibitem[Fauvaud et al.(2010)]{2010A&A...515A..39F} Fauvaud, S., Sareyan, J.-P., Ribas, I., et al.\ 2010, \aap, 515, A39.

\bibitem[Figer(1983)]{1983GEOCR...6.....F} Figer, A.\ 1983, GEOS Circular on RR Lyr Type Variables, 6


\bibitem[Fitch(1966)]{1966ApJ...143..852F} Fitch, W.~S.\ 1966, \apj, 143, 852.

\bibitem[Fitch \& Szeidl(1976)]{1976ApJ...203..616F} Fitch, W.~S. \& Szeidl, B.\ 1976, \apj, 203, 616.


\bibitem[Furgoni(2016)]{2016JAVSO..44....6F} Furgoni, R.\ 2016, Journal of the American Association of Variable Star Observers (JAAVSO), 44, 6

\bibitem[Garg et al.(2010)]{2010AJ....140..328G} Garg, A., Cook, K.~H., Nikolaev, S., et al.\ 2010, \aj, 140, 328.

\bibitem[Graham et al.(2019)]{2019PASP..131g8001G} Graham, M.~J., Kulkarni, S.~R., Bellm, E.~C., et al.\ 2019, \pasp, 131, 078001.

\bibitem[Handler et al.(1998)]{1998IBVS.4549....1H} Handler, G., Pikall, H., \& Diethelm, R.\ 1998, Information Bulletin on Variable Stars, 4549, 1

\bibitem[Hanson et al.(2004)]{2004AJ....128.1430H} Hanson, R.~B., Klemola, A.~R., Jones, B.~F., et al.\ 2004, \aj, 128, 1430.

\bibitem[Hoffmeister(1931)]{1931AN....242..129H} Hoffmeister, C.\ 1931, Astronomische Nachrichten, 242, 129.

\bibitem[Jayasinghe et al.(2020)]{2020MNRAS.493.4186J} Jayasinghe, T., Stanek, K.~Z., Kochanek, C.~S., et al.\ 2020, \mnras, 493, 4186.

\bibitem[Jeon et al.(2003)]{2003AJ....125.3165J} Jeon, Y.-B., Lee, M.~G., Kim, S.-L., et al.\ 2003, \aj, 125, 3165.

\bibitem[Jeon et al.(2004)]{2004AJ....128..287J} Jeon, Y.-B., Lee, M.~G., Kim, S.-L., et al.\ 2004, \aj, 128, 287.


\bibitem[Jeon et al.(2010)]{2010PASP..122...17J} Jeon, Y.-B., Kim, S.-L., \& Nemec, J.~M.\ 2010, \pasp, 122, 17.

\bibitem[Jurcsik et al.(2006)]{2006AA...445..617J} Jurcsik, J., Szeidl, B., V{\'a}radi, M., et al.\ 2006, \aap, 445, 617.

\bibitem[Kehoe et al.(2002)]{2002ApJ...577..845K} Kehoe, R., Akerlof, C., Balsano, R., et al.\ 2002, \apj, 577, 845.

\bibitem[Kochanek et al.(2017)]{2017PASP..129j4502K} Kochanek, C.~S., Shappee, B.~J., Stanek, K.~Z., et al.\ 2017, \pasp, 129, 104502.

\bibitem[Koch et al.(2010)]{2010ApJ...713L..79K} Koch, D.~G., Borucki, W.~J., Basri, G., et al.\ 2010, \apjl, 713, L79.

\bibitem[Kovacs \& Buchler(1994)]{1994AA...281..749K} Kovacs, G. \& Buchler, J.~R.\ 1994, \aap, 281, 749

\bibitem[Khruslov(2012)]{2012PZP....12....6K} Khruslov, A.~V.\ 2012, Peremennye Zvezdy Prilozhenie, 12, 6

\bibitem[Khruslov(2009)]{2009PZP.....9...26K} Khruslov, A.~V.\ 2009, Peremennye Zvezdy Prilozhenie, 9, 26

\bibitem[Khruslov \& Kusakin(2013)]{2013PZ.....33....6K} Khruslov, A.~V. \& Kusakin, A.~V.\ 2013, Peremennye Zvezdy, 33, 6

\bibitem[Khruslov et al.(2013)]{2013PZP....13...11K} Khruslov, A.~V., Huemmerich, S., \& Bernhard, K.\ 2013, Peremennye Zvezdy Prilozhenie, 13, 11

\bibitem[Khruslov(2007)]{2007PZP....7....25K} Khruslov, A.~V.\ 2007, Peremennye Zvezdy Prilozhenie, 7, 25

\bibitem[Khruslov(2011)]{2011PZP....11....30K} Khruslov, A.~V.\ 2011, Peremennye Zvezdy Prilozhenie, 11, 30

\bibitem[Khruslov \& Kusakin(2014)]{2014PZ.....34....1K} Khruslov, A.~V. \& Kusakin, A.~V.\ 2014, Peremennye Zvezdy, 34, 1

\bibitem[Khruslov(2014)]{2014PZP....14....1K} Khruslov, A.~V.\ 2014, Peremennye Zvezdy Prilozhenie, 14, 1

\bibitem[Lapukhin et al.(2015)]{2015PZP....15....8L} Lapukhin, E.~G., Veselkov, S.~A., \& Zubareva, A.~M.\ 2015, Peremennye Zvezdy Prilozhenie, 15, 8

\bibitem[Khruslov(2015)]{2015PZP....15....9K} Khruslov, A.~V.\ 2015, Peremennye Zvezdy Prilozhenie, 15, 9

\bibitem[Khruslov(2018)]{2018PZ.....38....1K} Khruslov, A.~V.\ 2018, Peremennye Zvezdy, 38, 1

\bibitem[Larson et al.(2003)]{2003DPS....35.3604L} Larson, S., Beshore, E., Hill, R., et al.\ 2003, BAAS

\bibitem[Lause(1932)]{1932AN....244..417L} Lause, F.\ 1932, Astronomische Nachrichten, 244, 417.


\bibitem[Li \& Qian(2010)]{2010AJ....139.2639L} Li, L.-J. \& Qian, S.-B.\ 2010, \aj, 139, 2639.

\bibitem[Mahra \& Sanyal(1968)]{1968Obs....88...58M} Mahra, H.~S. \& Sanyal, A.\ 1968, The Observatory, 88, 58

\bibitem[Mantegazza \& Poretti(1986)]{1986AA...158..389M} Mantegazza, L. \& Poretti, E.\ 1986, \aap, 158, 389

\bibitem[Mazur et al.(2003)]{2003MNRAS.340.1205M} Mazur, B., Krzemi{\'n}ski, W., \& Thompson, I.~B.\ 2003, \mnras, 340, 1205.

\bibitem[McNamara(2000)]{2000ASPC..210..373M} McNamara, D.~H.\ 2000, Delta Scuti and Related Stars, 210, 373

\bibitem[McNamara(2011)]{2011AJ....142..110M} McNamara, D.~H.\ 2011, \aj, 142, 110.

\bibitem[Mow et al.(2016)]{2016AJ....152...17M} Mow, B., Reinhart, E., Nhim, S., et al.\ 2016, \aj, 152, 17.

\bibitem[Musazzi et al.(1998)]{1998PASP..110.1156M} Musazzi, F., Poretti, E., Covino, S., et al.\ 1998, \pasp, 110, 1156.

\bibitem[Murphy et al.(2020)]{2020MNRAS.498.4272M} Murphy, S.~J., Saio, H., Takada-Hidai, M., et al.\ 2020, \mnras, 498, 4272.


\bibitem[Nemec(1989)]{1989upsf.conf..215N} Nemec, J.~M.\ 1989, IAU Colloq. 111: The Use of pulsating stars in fundamental problems of astronomy, 215

\bibitem[Nemec et al.(2017)]{2017MNRAS.466.1290N} Nemec, J.~M., Balona, L.~A., Murphy, S.~J., et al.\ 2017, \mnras, 466, 1290.

\bibitem[Netzel et al.(2021)]{2021arXiv210708064N} Netzel, H., Pietrukowicz, P., Soszy{\'n}ski, I., et al.\ 2021,

\bibitem[Niu et al.(2017)]{2017MNRAS.467.3122N} Niu, J.-S., Fu, J.-N., Li, Y., et al.\ 2017, \mnras, 467, 3122.

\bibitem[Niemczura et al.(2015)]{2015MNRAS.450.2764N} Niemczura, E., Murphy, S.~J., Smalley, B., et al.\ 2015, \mnras, 450, 2764.

\bibitem[Oosterhoff(1938)]{1938BAN.....8..277O} Oosterhoff, P.~T.\ 1938, \bain, 8, 277

\bibitem[Oosterhoff(1966)]{1966BAN....18..140O} Oosterhoff, P.~T.\ 1966, \bain, 18, 140

\bibitem[Palaversa et al.(2013)]{2013AJ....146..101P} Palaversa, L., Ivezi{\'c}, {\v{Z}}., Eyer, L., et al.\ 2013, \aj, 146, 101.

\bibitem[Petersen(1973)]{1973AA....27...89P} Petersen, J.~O.\ 1973, \aap, 27, 89

\bibitem[Petersen(1978)]{1978AA....62..205P} Petersen, J.~O.\ 1978, \aap, 62, 205

\bibitem[Petersen \& Christensen-Dalsgaard(1996)]{1996A&A...312..463P} Petersen, J.~O. \& Christensen-Dalsgaard, J.\ 1996, \aap, 312, 463

\bibitem[Pietrukowicz et al.(2013)]{2013AcA....63..379P} Pietrukowicz, P., Dziembowski, W.~A., Mr{\'o}z, P., et al.\ 2013, \actaa, 63, 379

\bibitem[Pietrukowicz et al.(2015)]{2015AcA....65...63P} Pietrukowicz, P., Latour, M., Angeloni, R., et al.\ 2015, \actaa, 65, 63

\bibitem[Pietrukowicz et al.(2020)]{2020AcA....70..241P} Pietrukowicz, P., Soszy{\'n}ski, I., Netzel, H., et al.\ 2020, \actaa, 70, 241.

\bibitem[Pigulski et al.(2006)]{2006MmSAI..77..223P} Pigulski, A., Ko{\l}aczkowski, Z., Ramza, T., et al.\ 2006, \memsai, 77, 223

\bibitem[P{\'o}cs \& Szeidl(2001)]{2001AA...368..880P} P{\'o}cs, M.~D. \& Szeidl, B.\ 2001, \aap, 368, 880. 

\bibitem[P{\'o}cs et al.(2002)]{2002AA...393..555P} P{\'o}cs, M.~D., Szeidl, B., \& Vir{\'a}ghalmy, G.\ 2002, \aap, 393, 555.

\bibitem[Poleski(2014)]{2014PASP..126..509P} Poleski, R.\ 2014, \pasp, 126, 509.

\bibitem[Pojmanski(2002)]{2002AcA....52..397P} Pojmanski, G.\ 2002, \actaa, 52, 397

\bibitem[Poretti(2003)]{2003A&A...409.1031P} Poretti, E.\ 2003, \aap, 409, 1031.

\bibitem[Poretti et al.(2005)]{2005AA...440.1097P} Poretti, E., Su{\'a}rez, J.~C., Niarchos, P.~G., et al.\ 2005, \aap, 440, 1097

\bibitem[Ramsay et al.(2011)]{2011MNRAS.417..400R} Ramsay, G., Napiwotzki, R., Barclay, T., et al.\ 2011, \mnras, 417, 400.

\bibitem[Ramsay et al.(2014)]{2014MNRAS.437..132R} Ramsay, G., Brooks, A., Hakala, P., et al.\ 2014, \mnras, 437, 132.

\bibitem[Rodr{\'\i}guez \& L{\'o}pez-Gonz{\'a}lez(2000)]{2000A&A...359..597R} Rodr{\'\i}guez, E. \& L{\'o}pez-Gonz{\'a}lez, M.~J.\ 2000, \aap, 359, 597

\bibitem[Shappee et al.(2014)]{2014AAS...22323603S} Shappee, B., Prieto, J., Stanek, K.~Z., et al.\ 2014, American Astronomical Society Meeting Abstracts \#223

\bibitem[Shi et al.(2021)]{2021PASP..133e4201S} Shi, X.-. dong ., Qian, S.-. bang ., Li, L.-. jia ., et al.\ 2021, \pasp, 133, 054201.

\bibitem[Skarka \& Caga{\v{s}}(2016)]{2016NewA...46...85S} Skarka, M. \& Caga{\v{s}}, P.\ 2016, \na, 46, 85.


\bibitem[\protect\citeauthoryear{{Stellingwerf}}{1979}]{1979ApJ...227..935S} {Stellingwerf}, R.~F. 1979, {\apj}, {227, 935}

\bibitem[Sterken et al.(2003)]{2003ASPC..292..121S} Sterken, C., Fu, J.~N., \& Brogt, E.\ 2003, Interplay of Periodic, Cyclic and Stochastic Variability in Selected Areas of the H-R Diagram, 292, 121


\bibitem[Su{\'a}rez et al.(2006)]{2006A&A...447..649S} Su{\'a}rez, J.~C., Garrido, R., \& Goupil, M.~J.\ 2006, \aap, 447, 649.

\bibitem[Su{\'a}rez et al.(2007)]{2007A&A...474..961S} Su{\'a}rez, J.~C., Garrido, R., \& Moya, A.\ 2007, \aap, 474, 961.

\bibitem[S{\"u}veges et al.(2012)]{2012MNRAS.424.2528S} S{\"u}veges, M., Sesar, B., V{\'a}radi, M., et al.\ 2012, \mnras, 424, 2528.


\bibitem[Templeton et al.(2002)]{2002ApJ...576..963T} Templeton, M., Basu, S., \& Demarque, P.\ 2002, \apj, 576, 963.

\bibitem[Templeton et al.(2009)]{2009PASP..121.1076T} Templeton, M.~R., Samolyk, G., Dvorak, S., et al.\ 2009, \pasp, 121, 1076.

\bibitem[van Cauteren \& Wils(2001)]{2001IBVS.5205....1V} van Cauteren, P. \& Wils, P.\ 2001, Information Bulletin on Variable Stars, 5205, 1

\bibitem[Walraven et al.(1992)]{1992MNRAS.254...59W} Walraven, T., Walraven, J., \& Balona, L.~A.\ 1992, \mnras, 254, 59.

\bibitem[Wang et al.(2014)]{2014NewA...32....6W} Wang, S.-M., Qian, S.-B., He, J.-J., et al.\ 2014, \na, 32, 6.

\bibitem[Watson et al.(2014)]{2014yCat....102027W} Watson, C., Henden, A.~A., \& Price, A.\ 2014, VizieR Online Data Catalog, B/vsx

\bibitem[Wils et al.(2003)]{2003IBVS.5442....1W} Wils, P., Lampens, P., Robertson, C.~W., et al.\ 2003, Information Bulletin on Variable Stars, 5442, 1

\bibitem[Wils et al.(2008)]{2008AA...478..865W} Wils, P., Rozakis, I., Kleidis, S., et al.\ 2008, \aap, 478, 865.


\bibitem[Wils et al.(2012)]{2012IBVS.6015....1W} Wils, P., Panagiotopoulos, K., van Wassenhove, J., et al.\ 2012, Information Bulletin on Variable Stars, 6015, 1


\bibitem[Wils et al.(2014)]{2014IBVS.6122....1W} Wils, P., Ayiomamitis, A., Robertson, C.~W., et al.\ 2014, Information Bulletin on Variable Stars, 6122, 1

\bibitem[Wils et al.(2015)]{2015IBVS.6150....1W} Wils, P., F-J, Hambsch., M. Vanleenhove et al., et al.\ 2015, Information Bulletin on Variable Stars, 6150, 1

\bibitem[Wo{\'z}niak et al.(2004)]{2004AJ....127.2436W} Wo{\'z}niak, P.~R., Vestrand, W.~T., Akerlof, C.~W., et al.\ 2004, \aj, 127, 2436.

\bibitem[Xue et al.(2018)]{2018ApJ...861...96X} Xue, H.-F., Fu, J.-N., Fox-Machado, L., et al.\ 2018, \apj, 861, 96

\bibitem[Yang et al.(2012)]{2012AJ....144...92Y} Yang, X.~H., Fu, J.~N., \& Zha, Q.\ 2012, \aj, 144, 92. 

\bibitem[Yang et al.(2018)]{2018RAA....18....2Y} Yang, T.-Z., Esamdin, A., Fu, J.-N., et al.\ 2018, Research in Astronomy and Astrophysics, 18, 002.

\bibitem[Yang et al.(2018)]{2018ApJ...863..195Y} Yang, T.-Z., Esamdin, A., Song, F.-F., et al.\ 2018, \apj, 863, 195

\bibitem[Yang \& Esamdin(2019)]{2019ApJ...879...59Y} Yang, T.-Z. \& Esamdin, A.\ 2019, \apj, 879, 59

\bibitem[Yang et al.(2021)]{2021AJ....161...27Y} Yang, T.-Z., Sun, X.-Y., Zuo, Z.-Y., et al.\ 2021, \aj, 161, 27.

\bibitem[Zong et al.(2019)]{2019PASP..131f4202Z} Zong, P., Esamdin, A., Fu, J.~N., et al.\ 2019, \pasp, 131, 064202.


\bibitem[Zorec \& Royer(2012)]{2012A&A...537A.120Z} Zorec, J. \& Royer, F.\ 2012, \aap, 537, A120.

\end{thebibliography}
\end{document}